# Thermal Switching of Lasing Regimes in Heavily Doped Er$^{3+}$ Fiber Lasers

Alexander M. Smirnov,*,† Alexey P. Bazakutsa,† Yuri K. Chamorovskiy,† Igor A. Nechepurenko,‡,§ Alexander V. Dorofeenko,‡,∥,⊥ Oleg V. Butov†

†Kotel'nikov Institute of Radioengineering and Electronics of RAS, Mokhovaya 11-7, Moscow 125009, Russia
‡Dukhov Research Institute of Automatics, 22 Suschevskaya, Moscow 127055, Russia
§National Research University Higher School of Economics, 20 Myasnitskaya, Moscow 101000, Russia
∥Institute for Theoretical and Applied Electromagnetics RAS, 13 Izhorskaya, Moscow 125412, Russia
⊥Moscow Institute of Physics and Technology, 19 Institutskiy pereulok, Dolgoprudny 141700, Russia

**ABSTRACT:** A pulsed regime of short-cavity, heavily erbium-doped fiber lasers is of high interest for its possible applications in telecommunications and sensorics. Here, we demonstrate these lasers in two configurations, distributed feedback laser and compare it with a classic Fabry-Perot type laser. We have managed to create lasers that function stably with cavities as small as 50 mm. Pulse properties such as amplitude, frequency and duration, are in a good agreement with our theoretical analysis, which takes into account spontaneous emission. We report the observation of the thermal switching effect, which consists of the pulsing regime changing to CW upon cooling the laser cavities down to the liquid nitrogen temperature. We theoretically show that this effect may be explained by weakening of the up-conversion process responsible for the pulsed regime. The slowing of the up-conversion processes is due to the energy mismatch in this process, which is overcome by interaction with phonons. At low temperatures, the number of phonons decreases and pulsing switches off.

Fiber lasers based on erbium-doped silica glass are actively used in modern fiber optics.[1-5] Erbium lasers operating at wavelengths near 1.5 μm are widely used in telecommunications,[6-10] optical sensing,[11-13] and radiophotonics.[14]

Of particular interest are single-frequency fiber laser systems with a narrow (about several kilohertz) radiation bandwidth,[15-18] which can serve as light sources for communication lines with ultra-high compaction of channels and coherent communication lines, as well as in sensor systems as a low-noise coherent source. Distributed feedback lasers (DFB lasers) are the most promising lasers in this aspect. A DFB fiber laser is a short (several centimeters) section of an active fiber into which a special Bragg grating with a phase shift is inscribed.[19-21] Due to the additional π phase shift introduced by the structure of the Bragg grating, the priority generation of only one longitudinal mode of the fiber laser is provided. As a result, in such lasers as narrow as 10 kHz bandwidth of the radiation spectrum is achieved at a low level of phase noise, which is especially important for high-precision sensing.

Miniaturization is one of the main goals of fiber laser development, which greatly expands the lasers' applications. Furthermore, obtaining a uniform-fiber Bragg structure of high length is quite a difficult technological challenge. Therefore, the ability to create a laser with a short cavity greatly simplifies the technology of its manufacture and stabilizes the laser oscillation. However, in order to attain an effective lasing in a short fiber section, the active fiber should possess both high gain and high absorption at a pumping wavelength. To meet these demands, achievement of a heavy-fiber core doping with rare earth metal ions at a minimum level of clustering is required. However, effective solubility of erbium ions in the fused silica glass has a limit, above which an intensive formation of ion clusters occurs. This establishes the maximum levels of gain and absorption for the pump. One widespread solution to this problem is the use of erbium fiber co-doped with ytterbium ions. This active fiber structure makes it possible to increase the pump efficiency at a low level of erbium doping and is currently actively used to create single-frequency lasers with short cavities.[22-25] However, this method has significant disadvantages. The most critical one is the high probability of spontaneous luminescence of ytterbium ions, which leads to amplified spontaneous emission (ASE)[26] and causes power drop and overall instability of the laser system.

It has previously been shown that the technology of surface plasma chemical vapor deposition of silica glass at reduced pressure (SPCVD)[27] for the preform's manufacture allows a significant increase in the level of the fiber core doping with rare-earth elements, while maintaining the minimum level of clustering due to the absence of the glass-melting stage during the synthesis process.[28, 29] Of course, at the stage of preform collapse, the process of fiber drawing is inevitably accompanied by a step of melting the formed structure. But investigations have shown that with the help of SPCVD technology, it is possible to increase rare earth ions' concentration in the silica glass of the fiber core without the formation of large cluster areas.[28] Previously, we have already demonstrated an ytterbium DFB laser with a cavity length of 12-16 mm,[24, 25] fabricated on the basis of heavily doped fiber synthesized by SPCVD technology. In the case of erbium doping, such a method will provide a sufficient level of pump absorption in a short cavity without co-doping the glass core with ytterbium. On the basis of the waveguide manufactured using this technology, a DFB laser with 24 mm cavity length was created, from which a stable laser oscillation was obtained.[25]

Erbium self-pulsed fiber lasers are of particular practical and scientific interest.[30-36] It was shown that[30, 37, 38] pulses occur due to the presence of $Er^{3+}$ ion pairs in the glass structure, and a dynamic laser model describing the self-pulsed and CW regimes in the presence of ion pairs in the system was proposed.

This paper presents the results of experimental and theoretical studies of lasing features using lasers with a 50 mm cavity length, whose structure was inscribed in heavily erbium-doped fibers. A pulsed generation mode was obtained in such lasers with a short duration (70 ns) and a high repetition rate (730 kHz). These are outstanding parameters, even compared to erbium lasers with saturable absorbers.[39-45] A modification of the theoretical model of $Er^{3+}$ laser operation was proposed, and the dependence of laser radiation intensity on time was obtained at different pump intensities. It is shown that these results are in good agreement with our experiment. An improved model and explanation of pulsed oscillation regime in $Er^{3+}$ lasers are proposed. To the best of our knowledge, the disappearance of the pulsed regime at liquid nitrogen temperatures has been demonstrated for the first time. An explanation of the observed effect is given herein.

On the basis a fiber made by us (see Materials and Methods, and absorption spectra of this fiber in Fig. 1), two erbium lasers with 50-mm-long cavities each were fabricated—namely, a distributed feedback erbium doped fiber laser (DFB) and a laser assembled according to the classical Fabry-Perot scheme with two Bragg mirrors at the cavity boundaries (FP-EDFL) (Fig. 2). The DFB laser has an infiber Bragg grating inscribed along the entire length of the cavity, having a phase shift ($\pi$-shift) in its structure displaced from the cavity center by about 10%. This displacement amount is optimal for obtaining unidirectional radiation generation of a DFB laser in single-mode regime with maximum power.[46] The measured transmission reduction at the Bragg wavelength ("spectrum depth") was approximately 21 dB. For the FP-EDFL laser, two Bragg gratings with lengths of 5 mm and 8 mm, respectively, were inscribed. The reflection coefficients of the output and high-reflection mirrors were 0.55 and 0.95, respectively. The identical cavity lengths of the two lasers allowed us to compare the operation of two different schemes.

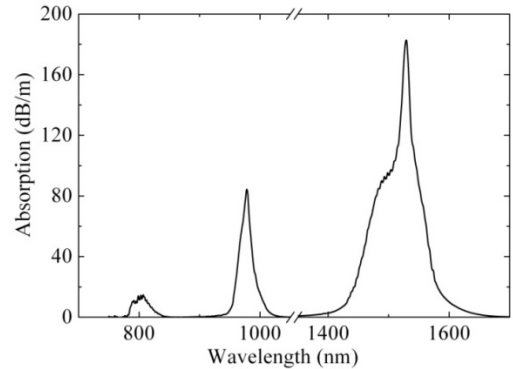

**Fig.1.** Absorption spectrum of erbium-doped fiber

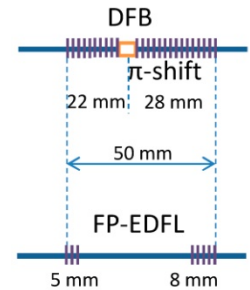

**Fig.2.** Scheme of investigated erbium fiber laser cavities.

Generation regimes of the investigated lasers were studied under the CW pump at a wavelength of 976 nm with the help of a Gooch & House semiconductor laser diode. The experimentally measured maximum output power of the pump laser radiation in fiber was 620 mW. To measure the spectra and time characteristics of the fiber laser's radiation, the experimental schemes shown in Fig. 3a and Fig. 3b were used. Laser cavities were installed in a cuvette with water or liquid nitrogen to stabilize the temperature.

The Bragg gratings of both lasers were inscribed directly into the active fiber cores by radiation of an ArF excimer laser (Coherent COMPex Pro) using a special 50-mm-long phase mask of 1075 nm period, which has the necessary phase shift in its structure. Let us note that in a number of papers, in order to provide photosensitivity, the erbium fiber was additionally co-doped with germanium. However, the presence of germanium in the erbium-doped fiber can lead to a decrease in gain,[28, 47, 48] which also has a detrimental effect on laser generation. In our samples, we did not use germanium. To inscribe the Bragg structures, we used a hydrogen-loaded fiber saturated with molecular hydrogen. It was shown earlier that erbium-doped fiber loaded *via* hydrogen allows us to inscribe the Bragg gratings in it with a sufficient efficiency even without germanium and phosphorus in the glass structure [25]. Hydrogen loading was carried out in a special chamber at a temperature of 70 °C and a pressure of 15 MPa for 48 hours.

The features of DFB laser operation have been investigated for various pumping schemes with different π-shift orientations relative to the pumping and radiation direction (see Results and Discussion for details), and also by using an additional fiber Bragg grating (FBG) with a Bragg wavelength of 976 nm. This grating returns the unabsorbed part of the pump radiation back to the cavity, thus increasing the efficiency of radiation absorption by the short laser cavity (Fig. 3a). In addition, a scheme using such a mirror makes it possible to distribute the pump energy more uniformly along the laser cavity, which is especially important in the case of the DFB laser, as it provides a more uniform heating of the laser cavity. A similar scheme with the additional reflecting Bragg grating at the pump wavelength was used in the case of FP-EDFL (Fig. 3b). A fiber coupler (50/50) was installed at the laser output for simultaneous measuring of the spectral and temporal characteristics of the laser radiation. Spectrum analysis was carried out using an optical spectrum analyzer (OSA) Agilent 86140B with a resolution of 0.06 nm. The Tektronix DPO4102B oscilloscope with a bandwidth of up to 1 GHz was used to measure the time characteristics.

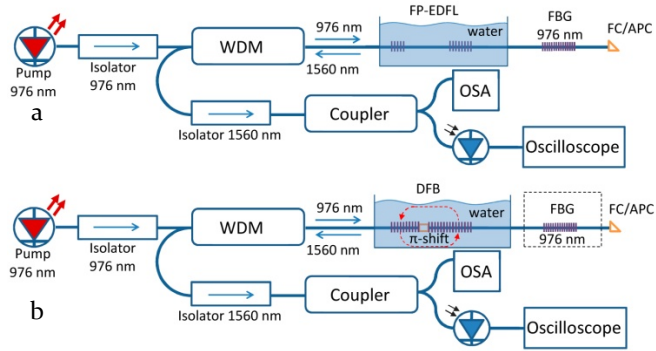

Fig.3. Experimental setups of the DFB (a) and the FP-EDFL (b) lasers. Here, OSA stands for the optical spectrum analyzer Agilent 86140B, WDM for the wavelength division multiplexer, and FBG for the fiber Bragg grating.

## RESULTS AND DISCUSSION

This section presents the results of an experimental study of the pulsed generation regimes of DFB and FP-EDFL lasers and the results of model calculations, as well as revealing and explaining the switching between the pulsed and CW regimes of laser operation when cooling lasers to the liquid nitrogen temperature.

**Pulsed generation regime: experiment**. The wavelength of the laser radiation was 1560 nm, as determined by the Bragg grating period. Both kinds of lasers operated in the self-pulsed regime, and their cavities were installed in a cuvette with water at room temperature. The duration and frequency of the pulses, as well as their intensity, depended on the scheme and on the pump power. A pulsing frequency of up to 730 kHz and peak power up to 0.7 W were achieved at maximum pumping power of 620 mW using the additional fiber Bragg grating.

When varying the pump power, the pulse duration was reduced from 2-3 μs observed at the lasing threshold to 69-71 ns at the maximum pump power. The dependence of the peak output power, pulse repetition rate and duration on the pump power are shown in Figs. 4, 5 and 6, respectively. As can be seen from these figures, both the DFB laser and classic FP-EDFL show quite similar behavior. When the DFB laser cavity was set in the inverse direction, we obtained a lower peak power, which is fully consistent with the concepts developed earlier.[34] These are associated with the phase shift position in the Bragg grating relative to the cavity center.

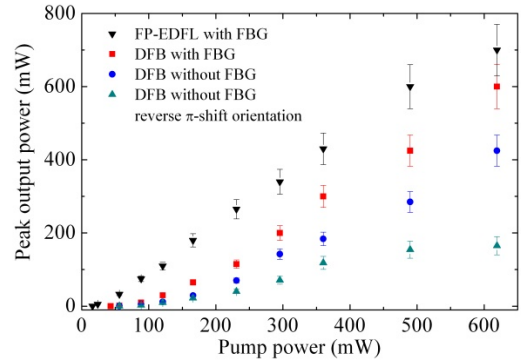

Fig. 4. The dependence of peak output power on pump power.

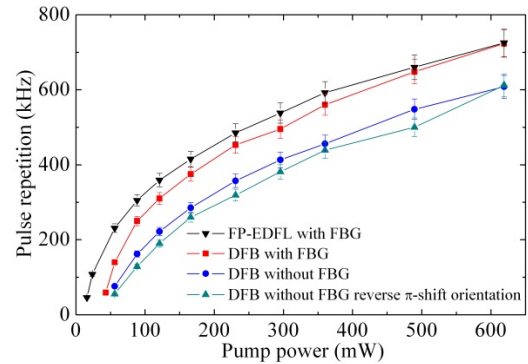

Fig. 5. The dependence of the pulse repetition frequency on pump power.

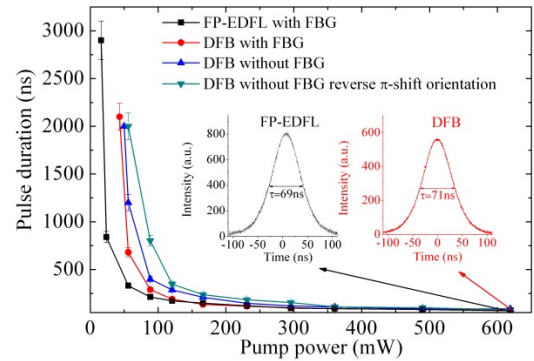

Fig. 6. The dependence of pulse duration on pump power.

It should be noted that, in general, the laser assembled according to the classical scheme (FP-EDFL) demonstrated a slightly better stability of the oscillation parameters than the DFB laser, which also manifested in a lower lasing threshold. A relatively high absorption level in the fiber leads to a significant and nonuniform heating of the laser cavity. This effect can be observed even when the fiber is placed in water, when the increase in the lasing wavelength associated with the heating of the Bragg grating is observed at the increase in pump power (Fig. 7). Since the absorbed power along the laser cavity is distributed inhomogeneously, this causes a chirp formation in the structure of the Bragg grating, which can have a negative effect on the parameters of the DFB cavity. As a result, the lasing threshold increases, and the reproducibility of pulse generation parameters deteriorates.

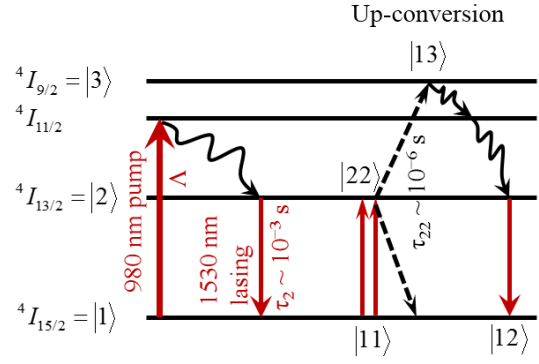

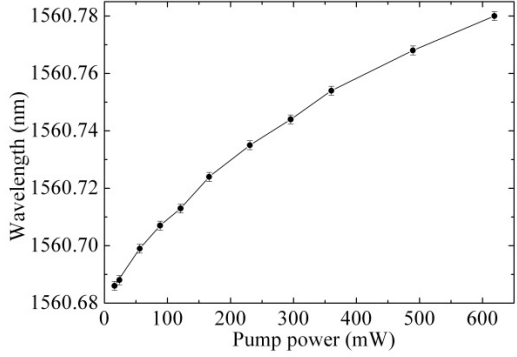

Fig. 7. The dependence of the lasing wavelength on the pump power of a DFB laser set in a cuvette with water.

**Pulsed generation regime: theoretical model.** Let us discuss the mechanism of pulse generation. A pulsed generation regime is typical of lasers in which erbium ions are used as an activator, especially at high erbium concentrations.[32, 33] A model explaining pulsed regime in such lasers was proposed elsewhere.[30, 37, 38] The reason for the appearance of pulses is the existence of ion pairs at high erbium concentrations. This creates an effective absorption dependent on levels' populations of ion pairs. This effect is interpreted as a saturable absorption.[30, 37, 38] It is well-known in laser physics that a saturable absorption causes passive Q-switching and a pulsed regime.[49, 50] From our point of view, passive Q-switching in highly doped $Er^{3+}$ lasers is realized due to this effect, which is opposite to the saturable absorption. The absorption grows with the field intensity instead of being saturated. To show this, let us consider the system in more detail.

Energy levels of $Er^{3+}$ are shown in Fig. 8. The main transition, at which 1.53 μm lasing occurs, is $^4I_{13/2} \to {^4I_{15/2}}$ (see left part of Fig. 8). The pumping wavelength is in the range of 976-980 nm, which corresponds to the $^4I_{15/2} \to {^4I_{11/2}}$ transition. The electron nonradiatively relaxes from $^4I_{11/2}$ to $^4I_{13/2}$ level, where the population inversion with respect to the $^4I_{15/2}$ level is created, resulting in the lasing.

Fig. 8. Energy level structure of a single $Er^{3+}$ ion in silica glass. In the left part of the figure, transitions of the $Er^{3+}$ ion in the lasing process are shown. In the right part of the figure, the up-conversion in ionic pair is shown. The straight and wavy solid lines show radiative and non-radiative relaxation processes, respectively. The dashed line shows a non-radiative energy transfer process between the ions in pair.

As noted above, pulse generation is related to the formation of $Er^{3+}$ ion pairs (miniclusters),[30, 31] in which up-conversion appears. The latter process can start when both ions occur in the first excited state, $^4I_{13/2}$. Then, one of the ions proceeds to the ground state, $^4I_{13/2} \to {^4I_{15/2}}$, with the non-radiative energy transfer to the second ion, $^4I_{13/2} \to {^4I_{9/2}}$. This process is possible due to the proximity of energies of the two $Er^{3+}$ ions in the $^4I_{13/2}$ state to energies of the two ions in $^4I_{15/2}$ and $^4I_{9/2}$ states, respectively (right part of Fig. 8). The excited ion non-radiatively relaxes to the $^4I_{13/2}$ level. The up-conversion rate is quite high—that is, 3 orders of magnitude higher than that of radiative transition $^4I_{13/2} \to {^4I_{15/2}}$.[30, 51] The system of rate equations used, which takes into account the up-conversion process, is derived from the one proposed earlier.[30] We use this system appended with a term responsible for the spontaneous emission process:[49]

$$\begin{aligned}
\frac{\partial d}{\partial t} &= \Lambda - a_2(1+d) - 2Id, \\
\frac{\partial d_+}{\partial t} &= a_2(1-d_+) - a_{22}(d_+ + d_-)/2 + yI(2-3d_+), \\
\frac{\partial d_-}{\partial t} &= \Lambda - a_2(1-d_+) - a_{22}(d_+ + d_-)/2 - yId_-, \\
\frac{\partial I}{\partial t} &= -I + A(1-2x)Id + AxyId_- + \xi.
\end{aligned} \quad (1)$$

Here, $d = n_2 - n_1$ is an average population inversion between the $^4I_{13/2}$ and $^4I_{15/2}$ states of a single ion; $d_\pm = n_{22} \pm n_{11}$ is a sum and difference of populations of the $|22\rangle$ ion pair state, with both ions being in the $^4I_{13/2}$ state, and of the $|11\rangle$ ion pair state, with both ions being in the $^4I_{15/2}$ state. The lasing intensity is described by variable $I$, the pumping rate by $\Lambda$, relaxation rates by $a_2 = \tau_l / \tau_2$ and $a_{22} = \tau_l / \tau_{22}$, which are the ratios of the electromagnetic mode relaxation time, $\tau_l$, to the relaxation time of single ion $|2\rangle$ state, $\tau_2$, and of the ion pair $|22\rangle$ state, $\tau_{22}$, respectively. The ratio of concentration of ion pairs to that of single ions is denoted by $x$; the ratio of the cavity field absorption cross-section by ion pair $\sigma'_l$ and by single ion $\sigma_l$ is denoted by $y = \sigma'_l / \sigma_l$. Parameter $A = \sigma_l N_0 \tau_l$ describes the strength of the field interaction with the gain medium. We have included a parameter $\xi$, which describes the spontaneous emission process. This parameter, which is used as a free parameter, substantially alters the shape of pulses near the lasing threshold, the latter being equal to:[37]

$$\Lambda_{th} = \frac{2a_2 a_{22}}{A} \frac{1+(1-2x)A + xyA}{2a_{22}(1-2x) + xy(2a_2 + a_{22})}. \quad (2)$$

Fig 9a shows oscillograms of pulsed oscillation for the example of DFB laser that were experimentally obtained at different pumping rates. Fig. 9b shows the corresponding curves $I(t)$ obtained by numerical solution of the system (1). Theoretical results are quite similar to the experimental ones. Differences between Figs. 9a and 9b are due to the simplicity of the model, which neglects a lot of effects, such as gain medium pumping inhomogeneity along the cavity, multimode effects in the long cavity, Stark splitting of levels, etc. We consider the obtained correspondence as highly acceptable with the account of all these idealizations.

The role of up-conversion processes is reduced to the appearance of losses, which abruptly increase when the ion pairs appear in the $|22\rangle$ state. To show this, let us write the equation for energy that follows from system (1):

$$\frac{\partial}{\partial t}\left[I + A(1-2x)d/2 + Axd_-\right] = -I + \xi \\ + A\Lambda/2 - a_2 A(1-2x)(1+d)/2 - a_2 A x n_{12} - a_{22} A x n_{22}. \quad (3)$$

The left-hand side of Eq. (3) is the rate of the total energy change of the "field + gain medium" system, whereas the right-hand side includes different channels of energy income and outcome. Particularly, the last term, $-a_{22} A x n_{22}$, is the rate of the energy dissipation due to the up-conversion process. This value is proportional to the small ion pair concentration $x$, and to the small population of the $|22\rangle$ state, $n_{22}$. On the other hand, the value of $a_{22}$ is much larger than $a_2$, $a_{22}/a_2 = \tau_2/\tau_{22} \sim 10^3$. This means that even a small $|22\rangle$ population leads to the appearance of large losses, which force the fast transition of the system to the $|12\rangle$ state during a period of some microseconds, whereas the duration of the radiative $|12\rangle \rightarrow |11\rangle$ relaxation is more of the order of 10 milliseconds. This means that instead of the saturable absorption, we deal with abrupt absorption growth with an increase in field intensity, the latter causing the ion pair excitation and fast energy dissipation. The pulsing mechanism in such a system becomes clear when observing the calculation results, which are shown in Fig. 10. First, let us note that almost all the ion pairs appear in the $|11\rangle$ and $|12\rangle$ states. Only a small fraction of ion pairs is excited to the $|22\rangle$ state, but these ion pairs provide considerable ab-sorption. When the population of the $|22\rangle$ state is lower than a certain critical value, the lasing starts. The field intensity grows with a characteristic time of nanoseconds, which follows from the assessment of the photon lifetime in the cavity, $\tau_l = L / \left[c(1 - r_1 - r_2)\right] \sim 0.3$ ns for the FP-EDFL laser, where $r_1$ and $r_2$ are the reflection coefficients of the two mirrors and $L$ is the cavity length. A lasing pulse develops and leads to the growth of the population of the $|22\rangle$ state of ion pairs, which leads to the increase in losses and breaks the lasing. The field decreases with the nanosecond characteristic time. In the absence of a field, the $|22\rangle$ level population drops with the microsecond characteristic time due to the up-conversion. Then, the process repeats. Therefore, ion pairs serve as a nonlinear, field-dependent dissipation source, which leads to the pulsed behavior.

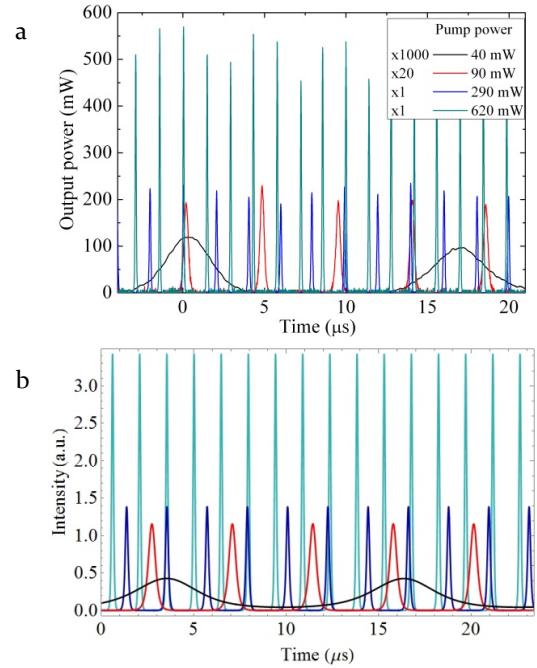

Fig. 9. Time dependence of lasing intensity of the DFB laser at different pumping rates obtained in the experiment (a) and in the calculations (b). For the experimental curves corresponding to 0.04 and 0.09 W, the oscillation peak power is shown with 1000x and 20x scaling, respectively. In the calculations, the pumping values are taken proportionally to the experimental ones, and the scaling coefficients are 50 and 5 for the black and red curves, respectively.

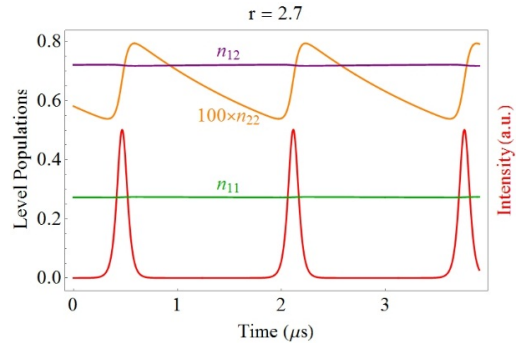

Fig. 10. Pulsing of the lasing intensity (red curve) and the populations of different ion pair states: $|11\rangle$ (both ions in the ground state), $|12\rangle$ (one of the ions is excited), $|22\rangle$ (both ions are excited).

**Switching to the CW oscillation regime with cooling.** It is worth noting that the pulsing frequency has the same order as that of relaxation oscillations;[52-56] that is, $10^5$ to $10^6$ Hz.[54] This correspondence is not a coincidence, because the increase in the pumping rate turns the laser from pulsed to CW generation.[30-32] This transition is performed through the increase in the pulsing rate, so that the pulses merge into a continuously oscillating curve. Then, pulses transform to relaxation oscillations, which decay over time, resulting in the CW lasing.

One can suppose that the change in the external conditions such as temperature can change the lasing regime at a constant pumping rate. We have investigated the behavior of erbium lasers under cooling by placing the cavities of our lasers into liquid nitrogen at 77 K. As a result, lasers of both types (FP-EDFL and DFB) switched to CW generation. In Fig. 11, lasing intensity as a function of the pumping rate is shown for the FP-EDFL. In the inset, one can observe oscillograms at different pumping rates, which confirms the CW regime.

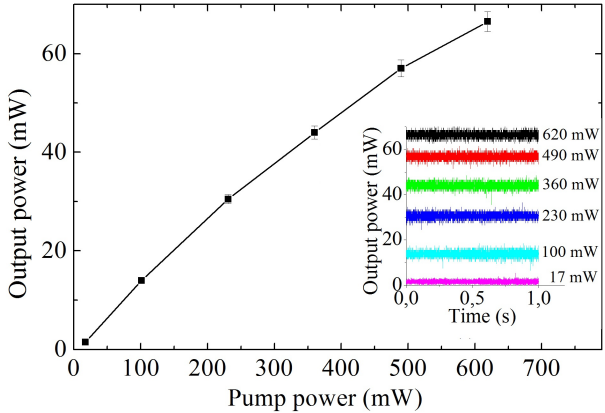

Fig. 11. FP-EDFL output power dependence on pump power at 77 K, confirming the CW generation regime.

The influence of temperature on the function of erbium lasers has been considered earlier in another regime, namely, under a coherent pumping. It was theoretically shown[31, 57] that the Stark sublevels' populations are determined by the Boltzmann distribution and, consequently, depend on temperature. According to the calculation results given in the cited papers, an increase in temperature should change the lasing regime from pulsing to CW. This effect is opposite from our observations. In [31], possible thermal change in the transitions' probabilities was not taken into account, which we consider crucial in our system with incoherent pumping.

Therefore, we built a model that describes the effect observed in our experiments. It is essential here that the frequencies of $|2\rangle \rightarrow |1\rangle$ and $|3\rangle \rightarrow |2\rangle$ transitions are quite proximate, but different. With the use of the data for the $|2\rangle \rightarrow |1\rangle$ and $|3\rangle \rightarrow |1\rangle$ wavelengths, $\lambda_{21}$=1550 nm and $\lambda_{31}$=810 nm,[58] we evaluate the frequency mismatch as $w = 2\pi c(2/\lambda_{21} - 1/\lambda_{31}) \sim 10^{14} s^{-1}$. In spite of this mismatch, the up-conversion process is possible because of the finite relaxation time. However, this time is not that of transition between any states (so-called longitudinal relaxation time, which is often denoted as $T_1$). Indeed, since the rate of the $|3\rangle \rightarrow |2\rangle$ transition is very high, the time spent in the intermediary state $|13\rangle$ is determined by the slower $|22\rangle$ transition, which has the same characteristic time as the whole up-conversion process, $10^{-6}$ s. This is several orders higher than $w^{-1} = 10^{-14} s$, and therefore cannot support a process with such high frequency mismatch. However, the linewidth is known[59] to be determined by another time, which is called a coherence time, or the transverse relaxation time, $T_2$. For most optical transitions, coherence time is of the order of $10^{-13} \div 10^{-15}$ s at room temperature, which corresponds well to the value of $w^{-1}$. The coherence time is mostly determined by the interaction with phonons and strongly depends on temperature. The coherence times of different transitions in $Er^{3+}$ and their temperature dependences were studied and measured in many papers; however, we could not find the coherence time of the up-conversion process in ion pairs. Therefore, for our estimation we use the coherence time of the $|2\rangle \rightarrow |1\rangle$ transition in a single $Er^{3+}$ ion, which has previously been measured.[60] At room temperature, coherence time is evaluated by the extrapolation of the thermal dependence to the value of $T_2 \sim 3 \times 10^{-13}$ s, whereas at 77 K the coherence time becomes an order of magnitude larger, $T_2 \sim 3 \times 10^{-12}$ s.[61]

Let us show that this increase in $T_2$ under cooling turns off the up-conversion process, which stops the pulse generation. It is shown in the Supplementary Material that the up-conversion rate, $\gamma_{22}$, is related to $T_2$ in the following way:

$$\gamma_{22} = \frac{2\Omega^2 T_2}{1 + w^2 T_2^2}, \quad (4)$$

where $\Omega$ is a parameter describing the interaction of ions in the pair (Rabi frequency). Eq. (4) gives a non-monotonic dependence of $\gamma_{22}$ on $T_2$ (Fig. 12). At small (below $w^{-1}$) values of $T_2$, one observes a growth of $\gamma_{22}$, which is due to the decrease in the transverse relaxation at the growth of $T_2$. Further growth of $T_2$ above $w^{-1}$ leads to the line narrowing below the frequency mismatch $w$, which leads to the decrease in $\gamma_{22}$.

It follows from the numerical values of $w$ and $T_2$ given above that at the room temperature we obtain the value $wT_2 \sim 30$, whereas at the liquid nitrogen temperature $wT_2 \sim 300$. Both points are located to the right from the point $wT_2 \sim 1$. Thus, growth of $T_2$ caused by cooling de-creases the up-conversion rate by approximately 10 times according to relation (4).

Switching between the lasing regimes caused by change in $T_2$ is illustrated by our "phase diagram" (Fig. 13), which is obtained by numerical solution of the system (1). At room temperature, pulsing is predicted in a wide range of pumping levels. At the liquid nitrogen temperature, only CW behavior should be observed. These results correspond exactly to our observations.

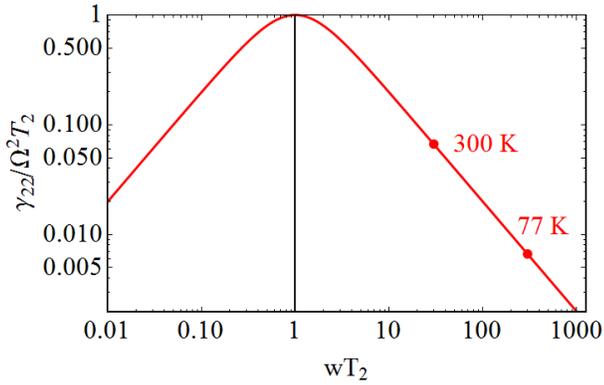

Fig. 12. Dependence of the up-conversion rate on coherence time. The points corresponding to the behavior of the system at room and liquid nitrogen temperatures are shown.

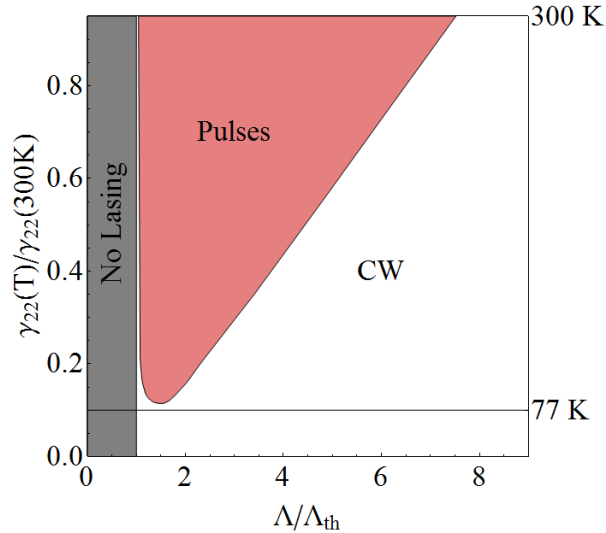

Fig. 13. The diagram of laser oscillation regimes in the coordinates of pump intensity ($\Lambda$) and up-conversion rate ($\gamma_{22}$). The red region corresponds to the pulsed regime, white region to the CW oscillation. Shown in gray is a sub-threshold region, where no oscillation is observed.

To conclude, our model has successfully described all the observed changes in the lasing regimes at the temperature change from the room to the liquid nitrogen temperatures. Substantially smaller changes in temperature lead to notable influence on the laser dynamics, as it follows from our numerical calculations. Therefore, erbium lasers are pulse sources that can be controlled by external conditions. Achieving higher doping levels and decreasing the cavity dimensions will lead to the expansion of application areas of fiber erbium lasers.

## MATERIALS AND METHODS

Active fiber was made from a preform with a core of silica glass doped with aluminum oxide and erbium oxide ($SiO_2$: $Al_2O_3/Er_2O_3$) and with an undoped silica cladding. The core glass was synthesized by means of SPCVD technology.[27, 28] It is necessary to note that the fiber core did not contain ytterbium ions. Vapor of anhydrous metal halides were used as initial reagents. Thanks to the SPCVD method, it was possible to produce fiber with a high content of erbium oxide (about 0.3 mol.%, which corresponds to $3.7 \times 10^{20}$ cm$^{-3}$) at a relatively low concentration of large cluster elements, which avoids a high level of "gray" losses in the active fiber. The absorption spectrum of the resulting fiber is shown in Fig. 1. The absorption level at the typical wavelengths of erbium transitions (1530 and 980 nm) was 180 dB/m and 90 dB/m, respectively.

## CONCLUSIONS

In this paper, we report on the observation of a new effect, namely, switching between pulsed and CW lasing regimes in heavily erbium-doped fiber lasers when the temperature decreases from the room temperature to liquid nitrogen values. It is theoretically shown that this effect is due to the energy levels' inexact match for the up-conversion process, which causes a significant role of temperature in its occurrence probability. To substantiate this assumption, we obtained estimates for the up-conversion rate and showed that this rate falls by about 10 times at the considered cooling of the system. A modified model of Er[3+] laser oscillation predicts the switching from self-pulsed to CW mode with such a decrease of the up-conversion process rate.

A detailed theoretical analysis of the nature of self-pulsed regimes in Er[3+] lasers was carried out. The main idea on the role of passive Q-switching in a self-pulsed regime was confirmed. However, our analysis showed that the appearance of pulses is not due to saturated absorption, but to the opposite effect, consisting of an increase in absorption with the growth of the oscillation intensity.

## ASSOCIATED CONTENT

**Supporting Information** provides a derivation of the relation (4) between the up-conversion rate and decoherence time

## AUTHOR INFORMATION

**Corresponding Author**
* email: alsmir1988@mail.ru

**Author Contributions**
†These authors are responsible for the experiment
‡These authors are responsible for the theory and numerical calculations

**Funding Sources**

This work was supported by the Russian Foundation for Basic Research (RFBR) (project 17-07-01388).

## ACKNOWLEDGMENT

We thank Prof. K.M. Golant for kindly providing the preform synthesized by SPCVD method.

## REFERENCES

(1) Bellemare, A., Continuous-wave silica-based erbium-doped fibre lasers. *Progress in Quantum Electronics* **2003,** *27* (4), 211-266.
(2) Bradley, J. D.; Pollnau, M., Erbium-doped integrated waveguide amplifiers and lasers. *Laser & Photonics Reviews* **2011,** *5* (3), 368-403.
(3) Brida, D.; Krauss, G.; Sell, A.; Leitenstorfer, A., Ultrabroadband Er: fiber lasers. *Laser & Photonics Reviews* **2014,** *8* (3), 409-428.
(4) Kringlebotn, J.; Archambault, J.-L.; Reekie, L.; Townsend, J.; Vienne, G.; Payne, D., Highly-efficient, low-noise grating-feedback $Er^{3+}$:$Yb^{3+}$ codoped fibre laser. *Electronics Letters* **1994,** *30* (12), 972-973.
(5) Weber, J. R.; Felten, J. J.; Cho, B.; Nordine, P. C., Glass fibres of pure and erbium-or neodymium-doped yttria–alumina compositions. *Nature* **1998,** *393* (6687), 769.
(6) Keller, U., Recent developments in compact ultrafast lasers. *Nature* **2003,** *424* (6950), 831.
(7) Wei, D.; Li, T.; Zhao, Y.; Jian, S., Multiwavelength erbium-doped fiber ring lasers with overlap-written fiber Bragg gratings. *Optics letters* **2000,** *25* (16), 1150-1152.
(8) Kim, S.; Park, J.; Han, S.; Kim, Y.-J.; Kim, S.-W., Coherent supercontinuum generation using Er-doped fiber laser of hybrid mode-locking. *Optics letters* **2014,** *39* (10), 2986-2989.
(9) Zhu, S.; Shi, L.; Xiao, B.; Zhang, X.; Fan, X., All-optical tunable microlaser based on an ultrahigh-Q erbium-doped hybrid microbottle cavity. *ACS Photonics* **2018**.
(10) Rönn, J.; Karvonen, L.; Kauppinen, C.; Perros, A. P.; Peyghambarian, N.; Lipsanen, H.; Säynätjoki, A.; Sun, Z., Atomic layer engineering of Er-ion distribution in highly doped Er: $Al_2O_3$ for photoluminescence enhancement. *ACS Photonics* **2016,** *3* (11), 2040-2048.
(11) Kim, H.; Kim, S.; Kim, B., Polarimetric fibre laser sensors using Er-doped fibre. *Optical and quantum electronics* **1995,** *27* (5), 481-485.
(12) Gong, Y.; Rao, Y.-J.; Guo, Y.; Ran, Z.-L.; Wu, Y., Temperature-insensitive micro Fabry–Perot strain sensor fabricated by chemically etching Er-doped fiber. *IEEE Photonics Technology Letters* **2009,** *21* (22), 1725-1727.
(13) Pérez-Herrera, R. A.; Quintela, M. A.; Fernández-Vallejo, M.; Quintela, A.; López-Amo, M.; López-Higuera, J. M., Stability comparison of two ring resonator structures for multiwavelength fiber lasers using highly doped Er-fibers. *Journal of Lightwave Technology* **2009,** *27* (14), 2563-2569.
(14) Tan, Y.-N.; Jin, L.; Cheng, L.; Quan, Z.; Li, M.; Guan, B.-O., Multi-octave tunable RF signal generation based on a dual-polarization fiber grating laser. *Optics express* **2012,** *20* (7), 6961-6967.
(15) Iwatsuki, K.; Okamura, H.; Saruwatari, M., Wavelength-tunable single-frequency and single-polarisation Er-doped fibre ring-laser with 1.4 kHz linewidth. *Electronics Letters* **1990,** *26* (24), 2033-2035.
(16) Ball, G.; Morey, W., Compression-tuned single-frequency Bragg grating fiber laser. *Optics letters* **1994,** *19* (23), 1979-1981.
(17) Mizrahi, V.; DiGiovanni, D. J.; Atkins, R. M.; Grubb, S. G.; Park, Y.-K.; Delavaux, J.-M., Stable single-mode erbium fiber-grating laser for digital communication. *Journal of Lightwave technology* **1993,** *11* (12), 2021-2025.
(18) Dong, X.; Ngo, N. Q.; Shum, P.; Tam, H.-Y.; Dong, X., Linear cavity erbium-doped fiber laser with over 100 nm tuning range. *Optics express* **2003,** *11* (14), 1689-1694.
(19) Groothoff, N.; Canning, J.; Ryan, T.; Lyytikainen, K.; Inglis, H., Distributed feedback photonic crystal fibre (DFB-PCF) laser. *Optics Express* **2005,** *13* (8), 2924-2930.
(20) Li, Q.; Yan, F.; Peng, W.; Feng, T.; Feng, S.; Tan, S.; Liu, P.; Ren, W., DFB laser based on single mode large effective area heavy concentration EDF. *Optics Express* **2012,** *20* (21), 23684-23689.
(21) Villanueva, G. E.; Pérez-Millán, P.; Palací, J.; Cruz, J. L.; Andres, M. V.; Marti, J., Dual-wavelength DFB erbium-doped fiber laser with tunable wavelength spacing. *IEEE Photonics Technology Letters* **2010,** *22* (4), 254-256.
(22) Kringlebotn, J.; Archambault, J.-L.; Reekie, L.; Payne, D., $Er^{3+}$:$Yb^{3+}$-codoped fiber distributed-feedback laser. *Optics Letters* **1994,** *19* (24), 2101-2103.
(23) Sejka, M.; Varming, P.; Hubner, J.; Kristensen, M., Distributed feedback $Er^{3+}$-doped fibre laser. *Electronics letters* **1995,** *31* (17), 1445-1446.
(24) Butov, O. V.; Rybaltovsky, A. A.; Bazakutsa, A. P.; Golant, K. M.; Vyatkin, M. Y.; Popov, S. M.; Chamorovskiy, Y. K., 1030 nm $Yb^{3+}$ distributed feedback short cavity silica-based fiber laser. *JOSA B* **2017,** *34* (3), A43-A48.
(25) Butov, O.; Rybaltovsky, A.; Vyatkin, M. Y.; Bazakutsa, A.; Popov, S.; Chamorovskiy, Y. K.; Golant, K. In *Short-cavity DFB fiber lasers*, Progress In Electromagnetics Research Symposium-Spring (PIERS), 2017, IEEE: 2017; pp 1594-1597.
(26) Han, Q.; Ning, J.; Sheng, Z., Numerical investigation of the ASE and power scaling of cladding-pumped Er–Yb codoped fiber amplifiers. *IEEE Journal of Quantum Electronics* **2010,** *46* (11), 1535-1541.
(27) Golant, K. In *Surface plasma chemical vapor deposition: 20 years of application in glass synthesis for lightguides (a review)*, Proceedings of XXI International Congress on Glass, 2007; p L13.
(28) Kholodkov, A.; Golant, K., $Er^{3+}$ ions luminescence in non-fused silicate glasses fabricated by SPCVD. *Optical materials* **2005,** *27* (6), 1178-1186.
(29) Kholodkov, A.; Golant, K. In *Surface plasma CVD as a new technological platform for Er-doped waveguide amplifiers and lasers fabrication*, Optical Fiber Communication Conference, 2004. OFC 2004, IEEE: 2004; p 3 pp. vol. 2.
(30) Sanchez, F.; Le Boudec, P.; François, P.-L.; Stephan, G., Effects of ion pairs on the dynamics of erbium-doped fiber lasers. *Physical Review A* **1993,** *48* (3), 2220.
(31) Loh, W., Suppression of self-pulsing behavior in erbium-doped fiber lasers with resonant pumping. *Optics letters* **1996,** *21* (10), 734-736.
(32) Li, N.; Bradley, J. D.; Singh, G.; Magden, E. S.; Sun, J.; Watts, M. R. In *Self-pulsing in Erbium-doped fiber laser*, Optoelectronics Global Conference (OGC), 2015, IEEE: 2015; pp 1-2.
(33) Hosseini, E. S.; Bradley, J. D.; Sun, J.; Leake, G.; Adam, T. N.; Coolbaugh, D. D.; Watts, M. R., CMOS-compatible 75 mW erbium-doped distributed feedback laser. *Optics letters* **2014,** *39* (11), 3106-3109.
(34) Skvortsov, M.; Wolf, A.; Dostovalov, A.; Vlasov, A.; Akulov, V.; Babin, S., Distributed feedback fiber laser based on a fiber Bragg grating inscribed using the femtosecond point-by-point technique. *Laser Physics Letters* **2018,** *15* (3), 035103.
(35) Fotiadi, A.; Kiyan, R.; Shakin, O., The effect of passive Q-switching observed in an erbium-doped fiber laser at a low pumping power. *Technical Physics Letters* **2001,** *27* (5), 434-436.
(36) Fotiadi, A. A.; Deparis, O.; Kiyan, R. V.; Chernikov, S.; Ikiades, A. In *Dynamics of passive Q-switching in SBS/Er fiber laser at low-pump-power*, Laser Optics 2000: Semiconductor

(37) Sanchez, F.; Stephan, G., General analysis of instabilities in erbium-doped fiber lasers. *Physical Review E* **1996,** *53* (3), 2110.

(38) Le Boudec, P.; Francois, P.; Delevaque, E.; Bayon, J.-F.; Sanchez, F.; Stephan, G., Influence of ion pairs on the dynamical behaviour of $Er^{3+}$-doped fibre lasers. *Optical and quantum electronics* **1993,** *25* (8), 501-507.

(39) Kurkov, A.; Sadovnikova, Y. E.; Marakulin, A.; Sholokhov, E., All fiber Er-Tm Q-switched laser. *Laser Physics Letters* **2010,** *7* (11), 795.

(40) Tsai, T.-Y.; Fang, Y.-C.; Hung, S.-H., Passively Q-switched erbium all-fiber lasers by use of thulium-doped saturable-absorber fibers. *Optics express* **2010,** *18* (10), 10049-10054.

(41) Luo, Z.; Huang, Y.; Zhong, M.; Li, Y.; Wu, J.; Xu, B.; Xu, H.; Cai, Z.; Peng, J.; Weng, J., 1-, 1.5-, and 2-μm fiber lasers Q-switched by a broadband few-layer $MoS_2$ saturable absorber. *Journal of Lightwave Technology* **2014,** *32* (24), 4077-4084.

(42) Huang, Y.; Luo, Z.; Li, Y.; Zhong, M.; Xu, B.; Che, K.; Xu, H.; Cai, Z.; Peng, J.; Weng, J., Widely-tunable, passively Q-switched erbium-doped fiber laser with few-layer $MoS_2$ saturable absorber. *Optics Express* **2014,** *22* (21), 25258-25266.

(43) Ren, J.; Wang, S.; Cheng, Z.; Yu, H.; Zhang, H.; Chen, Y.; Mei, L.; Wang, P., Passively Q-switched nanosecond erbium-doped fiber laser with $MoS_2$ saturable absorber. *Optics express* **2015,** *23* (5), 5607-5613.

(44) Kassani, S. H.; Khazaeinezhad, R.; Jeong, H.; Nazari, T.; Yeom, D.-I.; Oh, K., All-fiber Er-doped Q-switched laser based on tungsten disulfide saturable absorber. *Optical Materials Express* **2015,** *5* (2), 373-379.

(45) Zhang, M.; Hu, G.; Hu, G.; Howe, R.; Chen, L.; Zheng, Z.; Hasan, T., Yb-and Er-doped fiber laser Q-switched with an optically uniform, broadband $WS_2$ saturable absorber. *Scientific reports* **2015,** *5*, 17482.

(46) Loranger, S.; Tehranchi, A.; Winful, H.; Kashyap, R., Realization and optimization of phase-shifted distributed feedback fiber Bragg grating Raman lasers. *Optica* **2018,** *5* (3), 295-302.

(47) Barnes, W. L.; Laming, R. I.; Tarbox, E. J.; Morkel, P., Absorption and emission cross section of $Er^{3+}$ doped silica fibers. *IEEE Journal of Quantum Electronics* **1991,** *27* (4), 1004-1010.

(48) Furusawa, K.; Kogure, T.; Monro, T. M.; Richardson, D. J., High gain efficiency amplifier based on an erbium doped aluminosilicate holey fiber. *Optics Express* **2004,** *12* (15), 3452-3458.

(49) Siegman, A. E., Lasers university science books. *Mill Valley, CA* **1986,** *37*, 208.

(50) Svelto, O.; Hanna, D. C., *Principles of lasers*. Springer: 1998; Vol. 4.

(51) Delevaque, E.; Georges, T.; Monerie, M.; Lamouler, P.; Bayon, J.-F., Modeling of pair-induced quenching in erbium-doped silicate fibers. *IEEE Photonics Technology Letters* **1993,** *5* (1), 73-75.

(52) Ma, L.; Hu, Z.; Liang, X.; Meng, Z.; Hu, Y., Relaxation oscillation in $Er^{3+}$-doped and $Yb^{3+}/Er^{3+}$ co-doped fiber grating lasers. *Applied optics* **2010,** *49* (10), 1979-1985.

(53) Rønnekleiv, E., Frequency and intensity noise of single frequency fiber Bragg grating lasers. *Optical Fiber Technology* **2001,** *7* (3), 206-235.

(54) Turitsyn, S. K.; Babin, S. A.; Churkin, D. V.; Vatnik, I. D.; Nikulin, M.; Podivilov, E. V., Random distributed feedback fibre lasers. *Physics Reports* **2014,** *542* (2), 133-193.

(55) Ding, M.; Cheo, P. K., Analysis of Er-doped fiber laser stability by suppressing relaxation oscillation. *IEEE Photonics Technology Letters* **1996,** *8* (9), 1151-1153.

(56) Turitsyn, S. K.; Babin, S. A.; El-Taher, A. E.; Harper, P.; Churkin, D. V.; Kablukov, S. I.; Ania-Castañón, J. D.; Karalekas, V.; Podivilov, E. V., Random distributed feedback fibre laser. *Nature Photonics* **2010,** *4* (4), 231.

(57) Desurvire, E.; Simpson, J. R., Evaluation of $^4I_{15/2}$ and $^4I_{13/2}$ Stark-level energies in erbium-doped aluminosilicate glass fibers. *Optics letters* **1990,** *15* (10), 547-549.

(58) Kesavulu, C.; Sreedhar, V.; Jayasankar, C.; Jang, K.; Shin, D.-S.; Yi, S. S., Structural, thermal and spectroscopic properties of highly $Er^{3+}$-doped novel oxyfluoride glasses for photonic application. *Materials Research Bulletin* **2014,** *51*, 336-344.

(59) Pantell, R. H.; Puthoff, H. E., *Fundamentals of quantum electronics*. John Wiley & Sons: 1969.

(60) Zyskind, J.; Desurvire, E.; Sulhoff, J.; Di Giovanni, D., Determination of homogeneous linewidth by spectral gain hole-burning in an erbium-doped fiber amplifier with $GeO_2$:$SiO_2$ core. *IEEE Photonics technology letters* **1990,** *2* (12), 869-871.

(61) The line width at room temperature (300 K) is $\delta\lambda$ = 4 nm, which gives $T_2(300K) = \lambda^2/(2\pi c \cdot \delta\lambda) \sim 3\times10^{-13}$ s.[60] We also find[60] $1/T_2 \sim T^{1.61}$, with $T$ being temperature. Therefore, we have $T_2(77K)/T_2(300K) \sim 9$, so that $T_2(77K) \sim 3\times10^{-12}$ s.

Authors are required to submit a graphic entry for the Table of Contents (TOC) that, in conjunction with the manuscript title, should give the reader a representative idea of one of the following: A key structure, reaction, equation, concept, or theorem, etc., that is discussed in the manuscript. Consult the journal's Instructions for Authors for TOC graphic specifications.

Insert Table of Contents artwork here

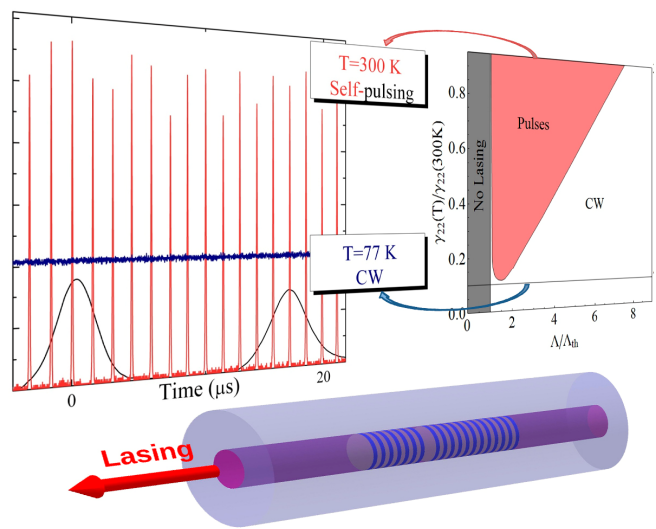